\documentclass[
    ,final            
  ]
  {aipproc}

\layoutstyle{8x11single}

\begin{document}

\title{$W$ and $Z$ Boson Production at $\sqrt s=1.96$ TeV with the D{\O} Detector}


\author{Alexey V. Popov for the D{\O} Collaboration}{
  address={Institute for High Energy Physics, Protvino}
}

\begin{abstract}
Preliminary measurements of the cross sections for the processes 
$p \bar p \rightarrow W \rightarrow \mu \nu, e \nu$; $p \bar p \rightarrow Z \rightarrow \mu \mu, e e$ 
using $0.1-0.3 fb^{-1}$ of Tevatron Run II data are presented. Measurement of the muon charge asymmetry from $W$ boson decays and $Z/ \gamma^* \rightarrow e e$ inclusive differential cross section as a function of boson rapidity are presented as well. The results are in agreement with Standard Model predictions \cite{csteor}.
\end{abstract}

\maketitle


\section{The D{\O} detector}

  The Run II D{\O} detector consists of the following main elements \cite{d0det}: central tracker, calorimeter and muon detector. The central tracker consists of a silicon microstrip tracker (SMT) and a central fiber tracker (CFT) which are both located within 2T superconducting solenoid magnet. The SMT was designed to optimize tracking and vertexing within $|\eta| < 3$. The system has a six barrel longitudinal structure interspersed with 16 radial disks. The CFT has eight coaxial barrels, each supporting two dublets of overlapping scintillator fibers. One doublet serves as axial and the other, alternating by $\pm 3^0$, serves as a stereo.
  
  The sampling calorimeter, made of uranium with liquid argon as active media, has a central section (CC) covering $|\eta| \leq 1 $ and two end caps (EC) covering $|\eta| < 4 $.
  
  The calorimeter is surrounded by the muon system consisting of three layers of scintillators and drift tubes with one layer inside 1.8T toroid and two layers outside. Muon identification at $|\eta| < 1$ relies on proportional drift tubes while tracking at $1 < |\eta| < 2$ relies on mini-drift tubes.
  
  Luminosity is measured using plastic scintillator arrays located in front of the EC cryostats covering $2.7 < |\eta| < 4.4$. The trigger system at D{\O} has three levels which reduce 1.7 MHz rate of inelastic collisions to 50 Hz that is written to tape.

\section{Cross sections for the $Z/W \rightarrow \mu \mu / \mu \nu$ processes}

\begin{figure}[h]
\begin{tabular}{lrrrr}
  \includegraphics[height=.25\textheight]{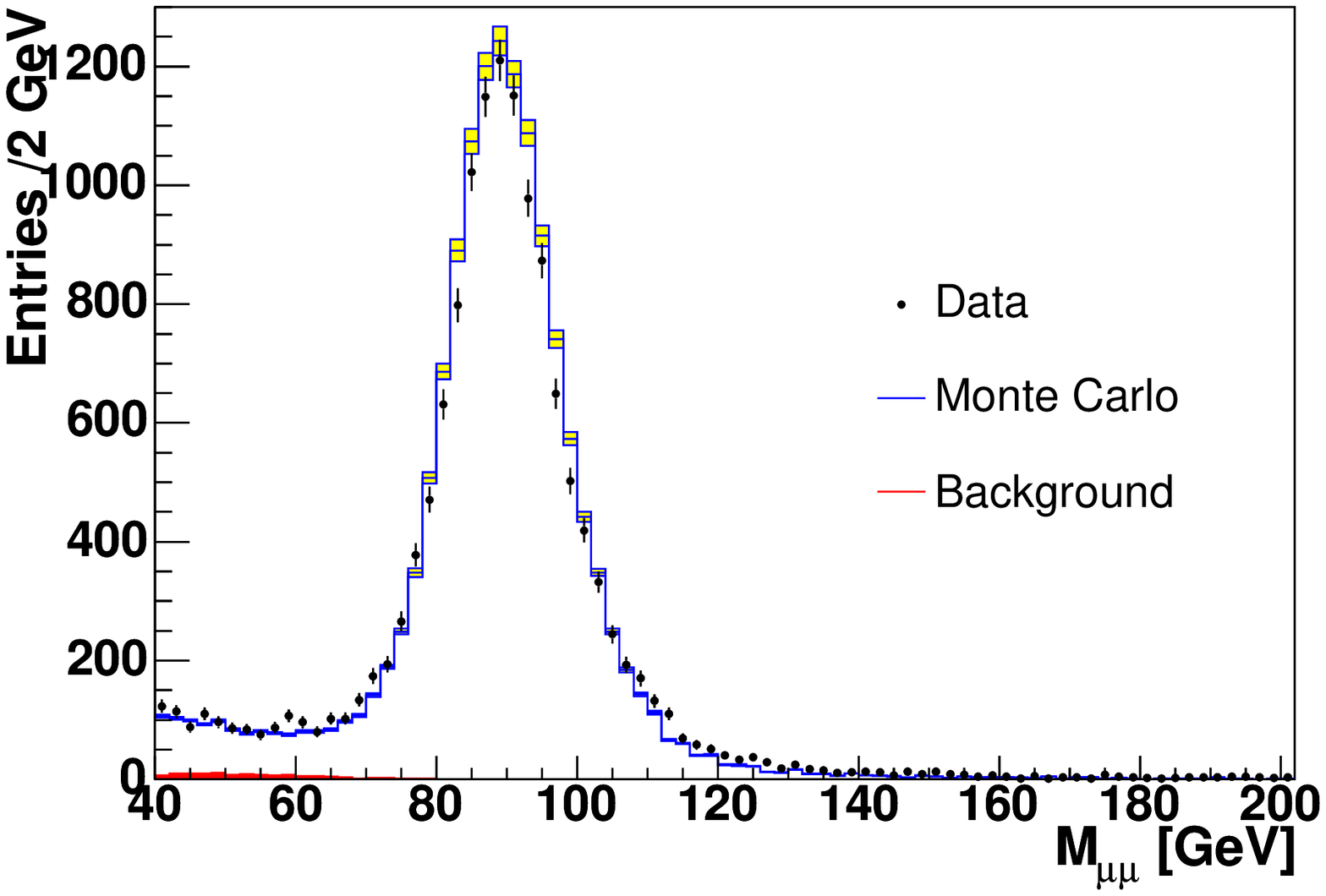} 
&
  \includegraphics[height=.25\textheight]{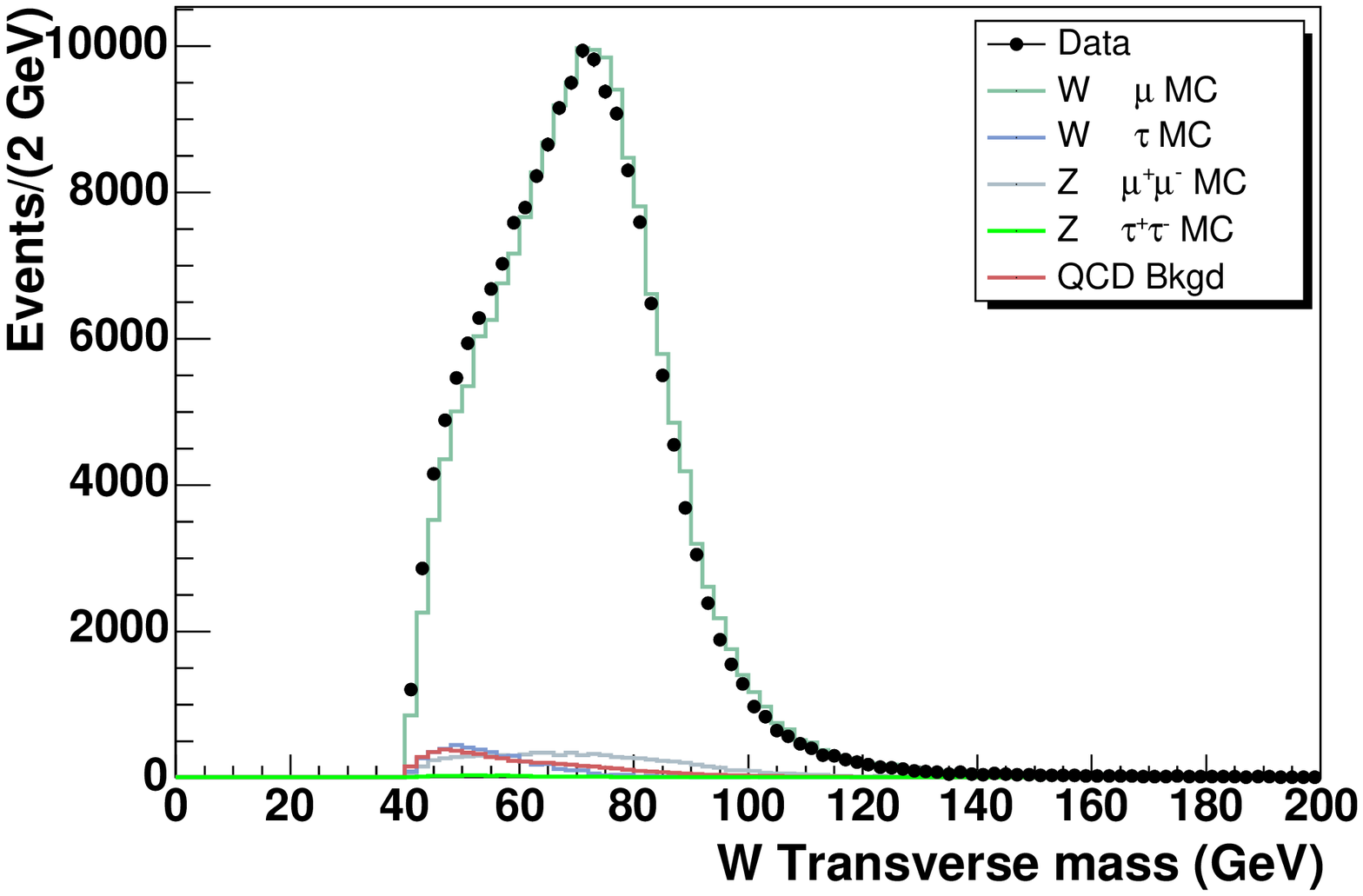} \\
\end{tabular}
\caption{{\bf Figure on the left}: comparison of $M_{\mu\mu}$ for $Z/\gamma^* \rightarrow \mu \mu$ events with
Monte-Carlo prediction. Shaded bands represent uncertainty on the Monte-Carlo due to the statistical unsertainties of the input efficiencies. Estimated background distribution is also shown on this plot. {\bf Figure on the right}: comparison of the $W$ transverse mass distribution for $W \rightarrow \mu \nu$ candidates and Monte-Carlo prediction of various backgrounds.}
\end{figure}

\subsection{Cross Section $\sigma(p\bar p \rightarrow Z X) \times Br(Z \rightarrow \mu \mu)$ }

Preliminary measurement of the cross section for the process $p \bar p \rightarrow Z/ \gamma^* \rightarrow \mu \mu$ in the mass range $M_{\mu\mu}>40$ GeV is performed \cite{zmumu} using the data sample corresponding to an integrated luminosity of $148 pb^{-1}$. A total of 14352 di-muon events are selected with an estimated background fraction $(0.5 \pm 0.3) \%$ arising from $b \bar b$, $(0.1 \pm 0.1) \%$ from cosmic rays, $(0.5 \pm 0.1) \%$ from $Z \rightarrow \tau \tau$ and $(0.2 \pm 0.1) \%$ from $W \rightarrow \mu \nu$ and di-boson backgrounds (see Figure 1). Measured cross section is:
\begin{equation}
\sigma (p \bar p \rightarrow Z/ \gamma^*) \times Br(Z/ \gamma^* \rightarrow \mu^+ \mu^-) = 329.2 \pm 3.4 (stat.) \pm 7.8 (syst.) \pm 21.4 (lumi.) pb
\end{equation}
Correcting the number of di-muon events by a factor of $0.885 \pm 0.015$ for the contribution from the pure photon exchange and $Z/ \gamma^*$ interference, the result 
\begin{equation}
\sigma (p \bar p \rightarrow Z) \times Br(Z \rightarrow \mu^+ \mu^-) = 291.3 \pm 3.0 (stat.) \pm 6.9 (syst.) \pm 18.9 (lumi.) pb
\end{equation}
is obtained.

\subsection{Cross Section $\sigma(p\bar p \rightarrow W X) \times Br(W \rightarrow \mu \nu)$}

A preliminary measurement of the cross section for the process $p \bar p \rightarrow W \rightarrow \mu \nu$ is performed \cite{wmunu} using the data sample corresponding to an integrated luminosity of $96 pb^{-1}$. A total of 62285 candidate events are observed, of which $7.8 \%$ are attributed to background processes (see Figure 1). Measured cross section is:
\begin{equation}
\sigma (p \bar p \rightarrow W) \times Br(W \rightarrow \mu \nu) = 2989 \pm 15 (stat.) \pm 81 (syst.) \pm 194 (lumi.) pb
\end{equation}

\section{Muon charge assymetry from the $W \rightarrow \mu \nu$ decay}

\begin{figure}[h]
  \includegraphics[height=.25\textheight]{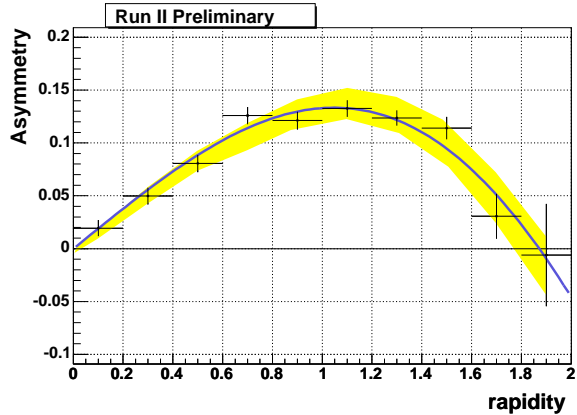}
  \caption{CP folded muon charge assymetry with combined statistical and systematic unsertainties. The yellow band is the total PDF unsertainty determined using the 40 CTEQ 6.1 PDF error sets. The blue line is expected assymetry using the MRST02 PDF. }
\end{figure}

A measurement of the muon charge assymetry from $W$ boson decays is performed using $\approx 230 pb^{-1}$ of Run II data \cite{wmunuass}. Measured  distribution is compared with expectation from NLO calculation using the CTEQ6.1M and the MRST02 parton distribution functions \cite{wmnpdf} and can be used as inputs to future PDF fits. Statistical uncertainties are greater than systematic unsertainties in every bin. This bodes well for the future of this analysis as more data, collected with the D{\O} detector, is analyzed. Measured assymetry distribution is presented on Figure 2.

\section{Cross-section for $W$ and $Z$ production in electron final states}

Preliminary measurements of the $W$ and $Z$ boson production cross sections times branching fractions into electrons using the data sample corresponding to an integrated luminosity of $177.3 pb^{-1}$ have been performed \cite{wze}. The measured cross sections are:
\begin{equation}
\sigma (p \bar p \rightarrow W) \times Br(W \rightarrow e \nu) = 2865.2 \pm 8.3 (stat.) \pm 76 (syst.) \pm 186.2 (lumi.) pb
\end{equation}
\begin{equation}
\sigma (p \bar p \rightarrow Z) \times Br(Z \rightarrow e^+ e^-) = 264.9 \pm 3.9 (stat.) \pm 9.9 (syst.) \pm 17.2 (lumi.) pb
\end{equation}
The ratio $R$ of the $W$ cross section times branching fraction to the $Z$ cross section times branching fraction is $R=10.82 \pm 0.16(stat.) \pm 0.28(syst.)$.

\section{Inclusive differential cross section for the $p \bar p \rightarrow Z/ \gamma^* \rightarrow e^+ e^-$ process as a function of boson rapidity}

\begin{figure}[h]
  \includegraphics[height=.25\textheight]{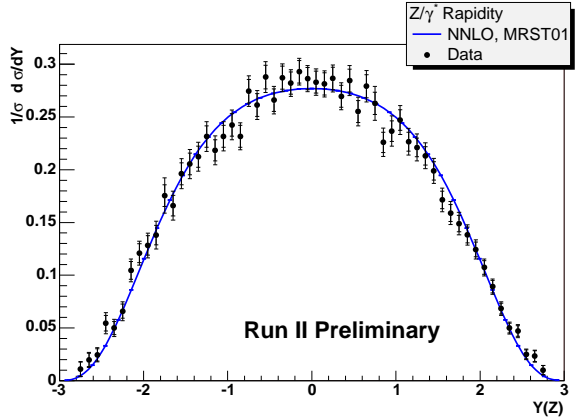}
  \caption{$d \sigma(p \bar p \rightarrow Z/ \gamma^* \rightarrow e e)/dY$ distribution. The outer error bars represent total error (combined statistical and systematic errors), while the inner error bars indicate the statistical error alone. The solid line shows the NNLO prediction based on the MRST 2001 PDFs.}
\end{figure}

The first Run II measurement of $Z/ \gamma^* \rightarrow e^+ e^-$ inclusive differential cross section as a function of boson rapidity in the mass range 71 to 111 GeV is performed using the data sample corresponding to an integrated luminosity of $337 pb^{-1}$\cite{zeerap}. At Run II Tevatron energy $Z$ bosons are produced with rapidities up to $\pm 3$. The cross section is measured over nearly the entire kinematic range. The data are in good agreement with the NNLO prediction based on the MRST 2001 parton distribution functions \cite{zeepdf} (Figure 3).

\end{document}